\documentclass[prb,twocolumn,superscriptaddress,showpacs,amsmath,amssymb]{revtex4-1}

\usepackage{graphicx}
\usepackage{latexsym}
\usepackage{amsmath}
\usepackage{amssymb}
\usepackage{amsfonts}
\usepackage{color}
\usepackage{bm}
\usepackage{verbatim}

\definecolor{MS-color}{RGB}{128,0,128}

\usepackage{framed}
\definecolor{shadecolor}{RGB}{222,222,221}
\begin{document}

\title{Generalized quasiclassical theory of the long-range proximity effect and spontaneous currents
in superconducting heterostructures with strong ferromagnets.}

 \date{\today}

\author{I. V. Bobkova}
\affiliation{Institute of Solid State Physics, Chernogolovka, Moscow
  reg., 142432 Russia}
\affiliation{Moscow Institute of Physics and Technology, Dolgoprudny, 141700 Russia}

\author{A. M. Bobkov}
\affiliation{Institute of Solid State Physics, Chernogolovka, Moscow reg., 142432 Russia}

 \author{M.A.~Silaev}
 \affiliation{Department of
Physics and Nanoscience Center, University of Jyv\"askyl\"a, P.O.
Box 35 (YFL), FI-40014 University of Jyv\"askyl\"a, Finland}


 \begin{abstract}
We present the generalized quasiclassical theory  of the long-range superconducting proximity effect
in heterostructures with strong ferromagnets, where the exchange splitting is of the order of Fermi energy.
In the ferromagnet the propagation of equal-spin Cooper pairs residing on the spin-split Fermi surfaces
is shown to be governed by the spin-dependent Abelian gauge field which results either
from the spin-orbital coupling or from the magnetic texture.
This additional gauge field enters into the quasiclassical equations
in superposition with the usual electromagnetic vector potential and results in the generation of spontaneous superconducting currents and phase shifts in various geometries which provide the sources of long-range spin-triplet correlations.
We derive the Usadel equations and boundary conditions for the strong ferromagnet
and consider several generic examples of the Josephson systems supporting spontaneous currents.
 \end{abstract}

  \maketitle

 \section{Introduction}

 Effective gauge theories has been introduced in many condensed matter systems
 including spin-triplet superfluid $^3$He \cite{VolovikBook}, cold atom systems \cite{Dalibard2011,Goldman2014} and magnetic
 materials \cite{Volovik1987,Nagaosa2013}.
  In spatially inhomogeneous magnetic textures the additional spin-dependent $U(1)$ gauge field of topological origin affects
 the motion of conduction electrons in the same way as the external electromagnetic field\cite{Volovik1987,Aharonov1992,Barnes2007,Hai2009,Yang2009}.
 That results in the topological Hall effect and emergent electrodynamics \cite{Bruno2004,Nagaosa2013} observed recently in the  chiral magnets
 with skyrmion lattices  \cite{Neubauer2009,Zang2011,Schulz2012,Liang2015}.

Geometric flux associated with the spin-dependent gauge field was predicted to generate  spontaneous
spin and charge currents in mesoscopic rings
with spatially-inhomogeneous texture of the Zeeman field \cite{Loss1990,Loss1992,Tatara2003,Tatara2003a}.
Up to now these effects have not been observed. Experimental  detection of persistent  currents in normal metals is in general rather
challenging \cite{Levy1990,Bleszynski-Jayich2009} since their magnitude is determined by the single-level
contribution which is rather small and highly sensitive to the details of disorder potential\cite{Buettiker1984,Cheung1989}.

 \begin{figure}[htb!]
 \centerline{$
 \begin{array}{c}
 \includegraphics[width=1.0\linewidth]{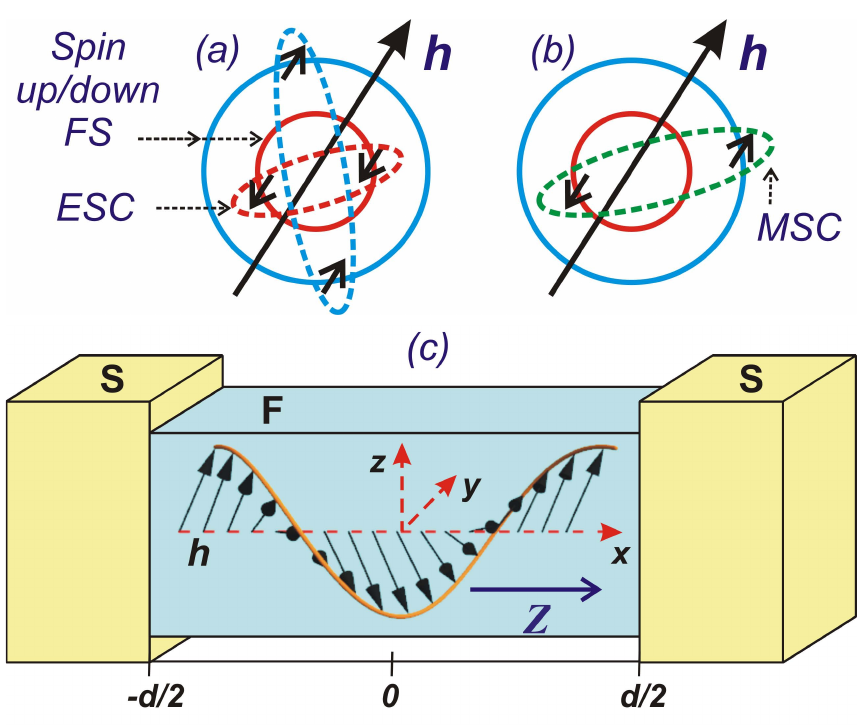}
   \end{array}$}
 \caption{\label{Fig:Model} (Color online)
 (a,b) Schematic picture of the Cooper pairs forming at the spin-split Fermi surfaces (FS).
  (a) The long-range equal-spin correlations (ESC) between the electrons
  within the spin-up or spin-down FS.
   (b) Mixed-spin correlations (MSC) that involve pairing between electrons in different spin-up/down FS.
   (c) Josephson junction through the strong ferromagnet (F) with magnetic helix texture
   producing the gauge field ${\bm Z} \parallel {\bm x} $ which generates spontaneous
   charge currents between superconducting (S) electrodes.
   }
 \end{figure}

The situation is completely different in the superconducting state
where the locally broken $U(1)$ gauge symmetry leads to the Meissner effect, i.e.
the generation of persistent condensate currents
in response to the external magnetic field.
However effects associated with the geometric spin-dependent flux  \cite{Loss1990,Loss1992,Tatara2003,Tatara2003a}
have not been identified in usual superconducting systems since the condensate of spin-singlet Cooper pairs is not sensitive
to the Zeeman field rotations.

 In the present paper we show that the superconducting condensate in fact can be coupled to the
  spin-dependent gauge fields emerging in superconductor/ferromagnet (SC/FM) hybrids.
 In such systems  the interplay of superconducting and magnetic orderings
 results in the generation of the spin-triplet correlations through the proximity effect\cite{Bergeret2005} .
 Of particular interest are the Cooper pairs with spin projections $S_z=\pm 1$
 in the local basis determined by the exchange field ${\bm h}$.
 They are formed by equal-spin correlations (ESC) between the electrons
 residing on one and the same spin-up/down Fermi surface [see Fig.(\ref{Fig:Model})a].
 Upon the adiabatic rotation of ${\bm h}$ such Cooper pairs
 pick up geometric phases and generate spontaneous superconducting currents.

 To understand the behaviour of ESC  we develop the
 gauge theory formalism to treat  proximity effect in SC/FM systems with spin-textured strong ferromagnets.
  So far, proximity and transport calculations in SC/FM hybrids have mostly concentrated
  on either fully polarized systems, so called half metals \cite{Eschrig2003,Braude2007,Eschrig2008, Mironov2015, Eschrig2015}, or
 in the opposite limit of weakly polarized systems \cite{Bergeret2005,Buzdin2005}, where the difference between
 spin subbands is completely neglected.
 However, most FMs have an intermediate exchange splitting of the energy bands of the order of
 but less than the Fermi energy.
 By now the quasiclassical theory for this regime has been formulated for the case of homogeneous magnetization of the strong
  ferromagnet \cite{Grein2009}.
 To describe the general situation we go beyond those limitations and consider
 SC/FM structures  with arbitrary large and spatially inhomogeneous exchange field.

 Our approach relies on the adiabatic approximation
 for spin transport\cite{Bliokh2005} which has been extensively used for studying transport phenomena
 in spin-textured magnets\cite{Volovik1987,Nagaosa2013}.
 We derive the quasiclassical equations describing ESC interacting with the spin-dependent $U(1)$ gauge field
 which can be induced either by the magnetic texture or spin-orbit coupling (SOC).
 The phases picked up by the $S_z=\pm 1$ Cooper pairs in response to this gauge field generate
 spontaneous superconducting currents through strong FMs.

 \section{Generalized quasiclassical theory}

  \subsection{The Model}

  We consider the Gor'kov equations in the presence of spin-dependent gradient terms,
  the exchange field ${\bm h}$ and the external vector potential ${\bm A}$:
  %
  %
  \begin{align} \label{Eq:Gorkov1}
  & ( \check G_0^{-1} + \mu - \check \Sigma ) \check G = \check I \delta ({\bm r} - {\bm r}^\prime)
  \\
  & \check G_0^{-1} =  -\hat\Pi \frac{1}{2m} \hat\Pi - \hat\sigma_k \{ M_{kj},  \hat p_j \}/2 +
  (i\omega + {\bm h} \bm{ \hat\sigma})\hat\tau_3 .
  \end{align}
 Here $\hat\Pi = \bm{ \hat p}- e {\bm A}\hat\tau_3$,
 $\check G = \check G ({\bm r},{\bm r}^\prime)$ is the matrix Green function (GF) in
 spin-Nambu space, $\{,\}$ is the anticommutator  added to have the hermitian Hamiltonian
  in the system with space-dependent field $ M_{kj}$,
  $\mu$ is the chemical potential, $\omega$ is the Matsubara frequency,
 $m$ is the effective mass which is equal to $m_F$ in the ferromagnet and to $m_S$ in the superconductor,
  $e$ is the electron charge, $\hat {\bm p} = (\hat p_x, \hat p_y, \hat p_z)$ is
 the momentum differential operator, $\hat\sigma_k$ and $\hat \tau_k$ are the spin and Nambu Pauli matrices. 
 The self-energy term $\check\Sigma$ includes the effects related to disorder scattering
 as well as the non-diagonal superconducting potential.
 The spin-dependent term $M_{kj}$ can be associated either with the SOC or with the pure gauge SU(2) field
 $
 M_{kj} = -i {\rm Tr} \left(\hat\sigma_k \hat U^\dagger \nabla_j \hat U\right) /2 m
 $
 where the transformation
 $ \hat U ({\bm r})= e^{i\bm {\hat\sigma }\bm{\theta}( {\bm r})/2} $
 rotates spin axes to the local frame where ${\bm h} \parallel {\bm z}$. It is  parametrized by the  spin vector $\bm \theta = \theta \bm n$ defined by the spatial texture of the exchange field
 distribution  $\bm h(\bm r)=\hat R( \bm \theta(\bm r))\bm h$, where $\hat R$ is the spatially-dependent rotation
 matrix and we choose ${\bm h}= h {\bm z}$. Therefore,  Eq.~(\ref{Eq:Gorkov1}) is written in the local reference frame, where the quantization axis is aligned with the local exchange field. We assume that the exchange field rotates slowly, on the large scales as compared to the atomic distances. For this reason we neglect second-order spatial derivatives of the exchange field. In the framework of this approach the inhomogeneity of the exchange field enters the equations as the pure gauge SU(2) field.

  In the general case of large  exchange splitting $|h| \sim \mu$
  the spin-dependent Gor'kov Eq. (\ref{Eq:Gorkov1}) is rather complicated and most importantly one cannot apply here the
  quasiclassical theory. The quasiclassical
  approximation is violated by the mixed-spin correlations (MSC) residing in spin-split
  subbands [Fig.~\ref{Fig:Model}b] which are characterized by the spatial length scale of the order of the Fermi wavelength
  $\lambda_F = \sqrt{2m/\mu}$.
  Therefore MSC yield vanishingly small contribution to the momentum-averaged observables at the distances much larger than the atomic length
  from the FM/SC interface. 
    Such correlations
  can be incorporated to the effective boundary conditions as the source terms for the
  ESC [Fig.~\ref{Fig:Model}a].  The ESC survive in the ferromagnet at much larger distances
  and can be treated within quasiclassics considered separately
  for each of the spin-up and spin-down Fermi surfaces.

  \subsection{Adiabatic approximation}

  To develop the quasiclassical approximation we divide the GF into the parts corresponding
  ESC of the spin-up/down states
  (see Fig.\ref{Fig:Model}a)
  \begin{equation}
  \check G_{ES} = \left(%
  \begin{array}{cc}
   G_0 + G_z \hat\sigma_z & F_x \hat\sigma_x + F_y \hat\sigma_y \\
  \tilde F_x \hat\sigma_x + \tilde F_y \hat\sigma_y &
  \tilde G_0 + \tilde G_z \hat\sigma_z \\
  \end{array}%
  \right)
  \end{equation}
  and the one corresponding to the MSC (see Fig.\ref{Fig:Model}b)
  \begin{equation}
  \check G_{MS} = \left(%
  \begin{array}{cc}
  G_x \hat\sigma_x + G_y \hat\sigma_y & F_0 + F_z \hat\sigma_z \\
  \tilde F_0 + \tilde F_z \hat\sigma_z &
  \tilde G_x \hat\sigma_x + \tilde G_y \hat\sigma_y \\
  \end{array}
  \right).
  \end{equation}
  Then from the Gor'kov equation (\ref{Eq:Gorkov1}) one can see that the amplitude of
  MSC is in general proportional to $\check{G}_{MS} \propto  ( M_{ij}/\lambda_F h) \check{G}_{ES}$
  where $\lambda_F$ is the Fermi wavelength. Therefore MSC are small as compared to that of the ESC
  if the adiabatic criterion is satisfied $| M_{ij}/\lambda_F h | \ll 1$.

  Within the adiabatic approximation neglecting the MSC in Eq.(\ref{Eq:Gorkov1}) and
  substituting the expansion of momentum operator
  $\hat p^2 = p^2 - 2i {\bm p} \bm\nabla $,
  we obtain the quasiclassical equation for ESC part:
  \begin{equation} \label{Eq:QuasiclassicsESQ}
  i\bm{\check V} \hat \nabla_{\bm R} \check G_{ES} +
  [ i\omega \hat\tau_3 - M_{zj} \hat\sigma_z p_j - \check\Sigma, \check G_{ES} ] =0 ,
  \end{equation}
 where $\hat \nabla_{\bm R} = \nabla_{\bm R} - i e {\bm A} [\hat\tau_3, \cdot]$ and
 $\bm{\check V} = \bm v_+ \check\gamma_+ + \bm v_- \check\gamma_-$.
 Here $v_\pm = \sqrt{2(\mu \pm h)/m_F}$ are the spin-dependent Fermi velocities
 determined on each of the spin-split Fermi surfaces labelled by the subscript $\sigma = \pm $.
 We introduce the projection operators to spin-up and spin-down states
 $\check\gamma_+  =  \hat\tau_\uparrow \hat\sigma_\uparrow + \hat\tau_\downarrow \hat\sigma_\downarrow  $
 and
 $\check\gamma_-  = \hat\tau_\uparrow \hat\sigma_\downarrow + \hat\tau_\downarrow \hat\sigma_\uparrow$, respectively,
 where $\hat\tau_{\uparrow (\downarrow)} = (\hat\tau_0 + (-) \hat\tau_z)/2$  and
 $\hat\sigma_{\uparrow (\downarrow)} = (\hat\sigma_0 + (-) \hat\sigma_z)/2$.

 The paired states on each of the Fermi surfaces are given by the
 corresponding parts of the equal-spin correlator:
 $\check G_\pm =  \check\gamma_\pm \check G_{ES} $.
 This decomposition allows to introduce quasiclassical propagators separately for
 spin-up and spin-down blocks
 \begin{equation}\label{Eq:DefinitionQuasiclassics}
  \hat g_{\sigma} =  -\oint \frac{d\xi_{p\sigma}}{\pi i} \check G_\sigma,
 \end{equation}
 where $\xi_{p\sigma} = p^2/2m_F + \sigma h - \mu$ and the notation $\oint$ means that the integration takes into account the poles of
 GF neat the corresponding Fermi surface.

  From Eq.(\ref{Eq:QuasiclassicsESQ}) we obtain  generalized Eilenberger
  equations fro the spin-less quasiclassical propagators
 \begin{equation}\label{Eq:Eilenberger}
  i {\bm v}_{\sigma}  \hat\partial_{\bm R} \hat g_{\sigma} +
  [ i\omega \hat\tau_3 - \hat\Sigma_{\sigma}, \hat g_{\sigma} ] =0,
  \end{equation}
  where the covariant operator is
  \begin{align} \label{Eq:DCovariant}
  \hat\partial_{\bm R} = \nabla_{\bm R} + i \sigma {\bm Z} [\hat\tau_3, \cdot] -
  i e {\bm A} [\hat\tau_3, \cdot]   \\
  \label{Eq:SU2Gauge}
  {\bm Z} = (M_{z x}, M_{z y}, M_{z z})m_F.
  \end{align}
 One can see that the Eilenberger-type equations for the spin-up/down correlations
 contain an additional U(1) gauge field ${\bm Z}$ which is added to the usual electromagnetic vector potential ${\bm A}$
 with the  opposite effective charges for spin-up and spin-down Cooper pairs.
 The U(1) field   is obtained by projecting the initial SU(2) field $\hat\sigma_k M_{ki}$
 to the basis of spin-triplet pairing states: ${\bm Z} =  - i (\hat U^{-1} \nabla \hat U )_{11} $.
 This reduction means that we neglect spin-flip transitions between the spin-up and spin-down Cooper pairs
 induced by the SU(2) potential. On a qualitative level it is equivalent to the adiabatic
 approximation in the single-particle problems that allows to describe the quantum system evolution in terms of the Berry
 gauge fields\cite{Bliokh2005}.

  Finally, the quasiclassical expression for the charge current is given by
  \begin{equation} \label{Eq:ChargeCurrent}
  {\bm j} = -\frac{i\pi T e}{2}   \sum_{\sigma=\pm} \sum_{\omega} \nu_{\sigma}
  \langle {\bm v}_{\sigma} {\rm Tr} [ \hat\tau_3\hat g_{\sigma}] \rangle ,
  \end{equation}
  where $\nu_{\sigma}$ are the spin-resolved DOS and $ \langle .. \rangle$
  denotes the averaging over the spin-split Fermi surface.

 \subsection{Usadel equation for ESC}
 Let us consider the system with large impurity scattering rate as compared to the
 superconducting energies determined by the bulk energy gap $\Delta$.
 In this experimentally relevant diffusive limit it is possible to derive the generalized
 Usadel theory with the help of the
 normalization condition $\hat g_{\sigma} ^2= 1$ which holds due to the commutator structure of the
 quasiclassical equations (\ref{Eq:DCovariant}).

 The impurity self-energy in the Born approximation is given by
 $\hat\Sigma_{\sigma} =
 \langle \hat g_{\sigma} \rangle/2i\tau_{\sigma}$.
 In the dirty limit we have
 \begin{equation} \label{Eq:Usadel}
 2\tau_{\sigma}
 ({\bm v}_{\sigma} \hat\partial_{\bm R}) \hat g_{\sigma} =
 - [ \langle \hat g_{\sigma} \rangle , \hat g_{\sigma} ].
 \end{equation}
The solution of Eq.~(\ref{Eq:Usadel}) can be found as $\hat g_{\sigma}= \langle \hat {g}_{\sigma} \rangle +\hat {\bm g}_{\sigma}^a \frac{\bm p_\sigma}{p_\sigma}$, where the anisotropic part of the solution $\hat {\bm g}_{\sigma}^a$ is small with respect to $ \langle \hat {g}_{\sigma} \rangle$. Making use of the relation $\{ \langle \hat g_{\sigma} \rangle , \hat {\bm g}_{\sigma}^a \} =0$, which follows from the normalization condition, one obtains
 \begin{equation}\label{Eq:UsadelExpansion}
 \hat {\bm g}_{\sigma}^a =
 - \tau_{\sigma}
 {\bm v}_{\sigma} \langle \hat g_{\sigma} \rangle
 \hat\partial_{\bm R} \langle \hat g_{\sigma} \rangle.
 \end{equation}

 Substituting to Eq.(\ref{Eq:Eilenberger}) and omitting the angle brackets we get the diffusion equation
 \begin{equation} \label{Eq:UsadelEquation}
 D_{\sigma} \hat\partial_{\bm R}(
 \hat g_{\sigma} \hat\partial_{\bm R} \hat g_{\sigma}) -
 [ \omega \hat\tau_3, \hat g_{\sigma} ] =0 ,
 \end{equation}
 where $D_{\sigma}$ are the spin-dependent diffusion coefficients, in the
 isotropic case given by  $D_{\sigma} = \tau_{\sigma}  v^2_{\sigma}/3$.
 The current is obtained substituting expansion (\ref{Eq:UsadelExpansion}) to the
 Eq.(\ref{Eq:ChargeCurrent})
  \begin{equation} \label{Eq:ChargeCurrentDiff}
  {\bm j} = \frac{i\pi T e}{2} \sum_{\sigma =\pm  } \sum_{\omega }
  \nu_\sigma D_{\sigma}
  {\rm Tr}[ \hat\tau_3 \hat g_{\sigma} \hat\partial_{\bm R}\hat g_{\sigma} ].
  \end{equation}

 Equations (\ref{Eq:UsadelEquation},\ref{Eq:ChargeCurrentDiff}) together with the boundary conditions derived in the next section
  provide the framework to study proximity effect-related phenomena in strong ferromagnets with large amount of disorder.
  This regime describes a different physical situation
  as compared to the previous works, where other approximations have been used.
  That concerns papers studying Rashba superconductor with large SOC\cite{Houzet2015} and that
  dealing with weak exchange fields and SOC within SU(2)-covariant formulation\cite{Bergeret2013,Bergeret2014,Konschelle2015,Tokatly2017}.
  The applicability condition of our theory requires that the exchange splitting has to be much larger than all other energy scales
  except the Fermi energy. In particular, in the bulk superconductor we have omitted the spin-flipping terms produced by various sources like
  SOC, magnetic texture or magnetic impurity scattering. However the spin-flipping terms are still important in strong ferromagnets/half metals  since they provide the source of ESC. Such correlations appear due to the conversion of spin-singlet Cooper pairs
  leaking from the superconductor electrode into the long-range spin-triplet ones.
  Below we study this conversion and boundary conditions for ESC using a generic model of SC/FM interface.  

 \section{Boundary conditions}
 The quasiclassical Eq.~(\ref{Eq:Eilenberger}) deals with the long-range ESC transformed adiabatically
 in space under the action of spin gauge fields. In FM/SC systems with usual spin-singlet superconductors such correlations can appear
 only in result of the non-adiabatic spin flip which converts MSC into the ESC \cite{Houzet2007,Eschrig2008,Eschrig2015}.
 This process occurs within the thin layer near FM/SC interface and can be described by the effective boundary conditions.
 To derive them we consider the simple microscopic model of the boundary with the non-magnetic potential barrier.
 The FM/SC interface is located at $x=0$, the normal ${\bm n} = n_x {\bm x}$
 directed from SC to FM and $n_x=\pm 1$. The momentum components ${\bm p}_\parallel$  in $yz$ plane
 parallel to the boundary are conserved. The effective masses $m_S$ and $m_F$ in the superconducting and ferromagnet regions
 are assumed to be different and the FM/SC boundary is characterized by the interfacial
 potential barrier of the strength $V\delta (x)$ added to the Hamiltonian $\hat G_0^{-1}$ in Eq.(\ref{Eq:Gorkov1}).

  Let us outline the general strategy to deriving the boundary conditions for quasiclassical ESC propagators.
  We need to solve the exact Gor'kov equations (\ref{Eq:Gorkov1}) near the boundary
  with accuracy up to the first order in SU(2) terms, which provide the conversion of MSC to ESC.
  Therefore we will use an expansion  by the small parameter $|M_{ij}p_j /h | \ll 1$.

 In result we will find the slow component of the anomalous function
  $\hat{ \tilde F} ( x,x^\prime) = (\check{G})_{21}$, where the index corresponds to the Nambu space. 
  This component does not contain fast oscillations as a function of the center of mass coordinate $X= (x+x^\prime)/2$.
  Let us denote such components as
 {
 $\hat {\tilde  {{\cal F }}} ( x, x^\prime) $ }. Below we will show that these correlations
 have the form
  \begin{equation} \label{Eq:SlowCorrelations}
  \hat{\tilde  {{\cal F }}}  =
  \left(%
  \begin{array}{cc}
  0 & {\tilde  {{\cal F }}}_{-} \\
  {\tilde  {{\cal F }}}_{+} &  0 \\
  \end{array}%
  \right) .
  \end{equation}
  The spin-up ${\tilde  {{\cal F }}}_{+}$ and spin-down ${\tilde  {{\cal F }}}_{-}$ pairing amplitudes are given by
  \begin{equation} \label{Eq:FSS}
  {\tilde  {{\cal F }}}_{\sigma}(x,x^\prime) =
  e^{-i n_x(p_{\sigma n} x-p_{\sigma n}^* x^\prime)} S_{\sigma \perp} K_{\sigma } ,
  \end{equation}
   where
  $p_{\sigma n}  = \sqrt{2m_F ( \mu + \sigma h ) - p_{||}^2 }$, and
  $S_{\sigma \perp} = ( M_{xi} + i\sigma M_{yi} )n_i / h $ is the combination of SU(2) field components that generate ESC
  in the ferromagnet  near the superconducting interface.
  Here the coefficient $K_\sigma$ incorporates the dependence of the pairing amplitude on the
  interface barrier strength, order parameter and effective masses.

 By writing Eq.(\ref{Eq:FSS}) we assume the averaging over the directions of the in-plane momentum. This is
  enough in the dirty limit although in the clean  case additional important effects resulting from
  the in-plane gradients of the exchange field ${\bm h}$ can be obtained beyond this approximation. Eq.(\ref{Eq:FSS}) is valid for $\omega >0$
  and for $\omega <0$ the amplitude  can  be obtained using symmetry relations, as discussed below.

  We need to find the GF near the FM/SC interface determined by the 1D Gorkov' equation (\ref{Eq:Gorkov1}) along $x$ coordinate.
   Since each GF  is the 2x2 matrix in spin space they can be represented in the form of two spinors
  \begin{align} \label{Eq:spinor}
  & \hat G = \left( \hat u_1 \;\;\hat u_2 \right)   \\
  & \hat {\tilde F} = \left( \hat v_1 \;\;\hat v_2 \right),
  \end{align}
  where the spinor elements are
  $\hat u_k = (u_{k1}, u_{k2})^T$ and
  $\hat v_k = (v_{k1}, v_{k2})^T$.
   Let us consider the components $\hat u_1$, $\hat v_1$ in detail.
 The other pair $\hat u_2$, $\hat v_2$ is given by  $\hat u_2 =\sigma_x \hat u_1$ and $\hat v_2 = \sigma_x \hat v_1$
 and changing the sign of the fields $h$ and $M_{yi}$.

 For the Nambu spinor $\hat \psi = (\hat u_1, \hat v_1 )^T$ from (\ref{Eq:Gorkov1}) we obtain the equation
 in the ferromagnet
 \begin{equation}
   \left[ \mu_F - \frac{\hat p_x^2}{2m_F} -  \sigma_k \{  M_{kj}, \hat p_j  \}{ /2}   +
   \hat\tau_3 (i\omega  + h\sigma_z ) \right] \hat\psi = 0 \\
 \end{equation}
 and in the superconductor
  \begin{equation}
  \left[ \mu_S - \frac{\hat p_x^2}{2m_S} { +}
   \hat\tau_3 (i\omega + h\hat\sigma_z) \right] \hat\psi = 0  ,
 \end{equation}
 where  $\mu_{S,F} = \mu - p_\parallel^2/2m_{S,F}$ .
These equations look similar to the Bogolubov-de Gennes equation but taken at the imaginary frequency.
 The boundary condition are obtained by integrating Eq.(\ref{Eq:Gorkov1})
 near the singularity points $x=x^\prime$ and $x=0$.
  In this way we obtain the boundary conditions at $x=x^\prime$:
   \begin{align}  \label{Eq:bc2-1-1}
  & \hat\psi (x^\prime +0) =  \hat\psi (x^\prime -0) \\
   \label{Eq:bc2-1}
  & \nabla_x \hat u_1 (x^\prime +0) - \nabla_x \hat u_1 (x^\prime -0) = 2 m_F (1, 0)^T
  \\ \label{Eq:bc2-2}
  & \nabla_x \hat v_1 (x^\prime +0) - \nabla_x \hat v_1 (x^\prime -0) =0 ,
   \end{align}
  ant at $x=0$
  \begin{align} \label{Eq:bc1-1}
  & \hat\psi (+ 0 ) =  \hat\psi ( - 0 )
  \\ \label{Eq:bc1-3}
  & \frac{\nabla_x \hat\psi (- 0 ) }{m_F} - \frac{\nabla_x \hat\psi (+ 0)}{m_S}
  = ( { i M_{kx}\hat\sigma_k } - 2 V  ) \hat \psi (0).
  \end{align}

 Here we neglect the impurity scattering and the inverse proximity effect in the superconducting region.
 The impurity self energy can be included into the consideration in case of the
 tunnelling limit when the surface barrier is strong enough to suppress the inverse proximity effect in the superconductor.
 This consideration demonstrates that the impurity scattering does not change boundary conditions for quasiclassical
 propagators. In the opposite case of weak barriers one should strictly speaking take into account how the impurity
 self energy in the superconductor is modified by the inverse proximity effect, which in principle can affect the boundary conditions.

 Let's assume that ${n_x<0}$. Then the solution for electron wave in bulk normal ferromagnet can be written as
 $\hat v =0$ and
  \begin{align} \label{Eq:u1Incidend}
  & \hat u(x>x^\prime) =
    e^{ip_{+ n}x} \hat A_{+ >} +
   e^{ip_{-n}x} \hat A_{- >}
   \\
  & \hat u(x<x^\prime) =
   e^{-ip_{+n}x} \hat A_{+ <} +
   e^{-ip_{-n}x} \hat A_{- <} ,
   \end{align}
   where the labels $> (<)$ denote right- and left-going waves,
   $p_{\sigma n} = \sqrt{2m_F ( \mu_F +\sigma h + i\omega) }$ and we neglect SO corrections to the momenta.
   Expressions (\ref{Eq:u1Incidend}) are valid for the half-metal when $h > \mu_F $ as well. In this case one should take
   $p_{-n} = i \sqrt{2m_F (  h - \mu_F) }$.

     To the first order by the small parameter $|M_{ij} p_j/h | \ll 1$ we get expressions for
     the amplitudes in Eq.(\ref{Eq:u1Incidend})
     \begin{align} \label{Eq:A-up-plus}
     & \hat A_{+ >} =
     a^{(0)}_{+ >} e^{-ip_{+n} x^\prime}
     \left[ 1, \; - ( p_{+n} S_{\perp + }  + p_y S_{\parallel + })  \right]^T
     \\
     & \hat A_{- >} =
     a_{- >} e^{-ip_{+n} x^\prime}
     \left[( p_{-n} S_{\perp -} + p_y S_{\parallel -}), \; 1 \right]^T
     \\
     \label{Eq:A-up-minus}
     & \hat A_{+ <} =
     a^{(0)}_{+ <} e^{ip_{+n} x^\prime}
     \left[ 1, \; (p_{+n} S_{\perp +} - p_y S_{\parallel +}) \right]^T
     \\
     & \hat A_{- <} =
     a_{- <} e^{ip_{+n} x^\prime}
     \left[ ( p_y S_{\parallel -} - p_{-n} S_{\perp -}), \; 1 \right]^T ,
     \end{align}
     where $S_{\sigma \parallel} = ( M_{xy} + i\sigma M_{yy} )/ h $ and $S_{\sigma \perp}$
     is defined above.
      The zero - order amplitudes are given by
     \begin{align}
     a^{(0)}_{+ >} = a^{(0)}_{+ <}  =  -i m_F / p_{+n} .
     \end{align}
     We neglect the second term in Eq.(\ref{Eq:u1Incidend}) $ \hat A_{- >} $ since its amplitude is much smaller
     than $a^{(0)}_{+ >}$ due to the prefactors $p_{\sigma n} S_{\sigma \perp }$ and
     $p_y S_{\sigma \parallel }$.
     Therefore up to the first order in these small parameters
      we should take into account the reflected holes
     generated by  $ \hat A_{- >} $ without spin flip which have the same wave vector as the incident wave and therefore
     does not contribute to the slow-varying correlation (\ref{Eq:SlowCorrelations}).

   The solution (\ref{Eq:u1Incidend}) can be considered as the incident electronic wave
   at the FM/SC interface. The reflected wave consists of electronic $\hat u_r$ and hole $\hat v_r$ components
   having the form
   \begin{align} \label{Eq:ur}
   \hat u_r (x,x^\prime) =  e^{-ip_{+n} x} \hat D_{+} (x^\prime) +
   e^{-ip_{-n} x} \hat D_{-} (x^\prime) \\
   \label{Eq:vr}
   \hat v_r (x,x^\prime) = e^{ip^*_{-n} x} \hat B_{+} (x^\prime) +
    e^{ip^*_{+n} x} \hat B_{-} (x^\prime) .
   \end{align}
   Here the structure of $\hat D_{\sigma}$, is similar to that of $ \hat A_{\sigma <}$
    \begin{align}  \label{Eq:D}
   & \hat D_{+}(x^\prime) = d_+ e^{-ip_{+n} x^\prime}
   \left[ 1, \;
    (p_{+n} S_{\perp +} - p_y S_{\parallel +})
      \right]^T
   \\    \label{Eq:Dm}
   & \hat D_{-}(x^\prime) = d_{-} e^{-ip_{+n} x^\prime}
   \left[
    ( p_y S_{\parallel -} - p_{-n} S_{\perp -}), \;
    1 \right]^T .
   \end{align}
    The reflected hole-like wave is given by (\ref{Eq:vr}) with the amplitudes
    \begin{align} \label{Eq:B}
   & \hat B_+ (x^\prime)  = b_+ e^{-ip_{+n} x^\prime}
    \left[ 1, \;
     (S_{\perp +} p^*_{-n} + S_{\parallel +} p_y )
     \right]^T
    \\ \label{Eq:Bm}
   & \hat B_- (x^\prime) = b_- e^{-ip_{+n} x^\prime}
   \left[
    -( S_{\perp -} p_{+n}^* + S_{\parallel -} p_y ) , \;
    1
    \right]^T .
   \end{align}

   We are interested  in the wave $\hat B_-$ because the corresponding contribution
   to the reflected hole amplitude $\hat v_r$ given by the second term in Eq.(\ref{Eq:vr}) does not contain fast
   oscillations as the function of $X=(x+x^\prime)/2$. Hence it provides the source of the long-range
   superconducting correlations.
   On a qualitative level the wave $\hat B_- $ determines the {\it spin-flip Andreev reflection}
   leading to the generation of spin-triplet Cooper pairs.

   Hence we obtain the component of the ESC in the form
   ${{\tilde  {{\cal F }}}_{+}(x,x^\prime) = b_- e^{i (p_{+n } x-p_{+n}^* x^\prime)}} $ .
   The other component is obtained from the other pair of spinors $(\hat u_2, \hat v_2)$
   and has the form ${{\tilde  {{\cal F }}}_{-}(x,x^\prime) = b_+ e^{i (p_{- n} x-p_{- n}^* x^\prime)} }$.
   Averaging over in-plane momentum directions we get ${\langle b_{\bar \sigma} \rangle = S_{\sigma \perp} K_{\sigma}}$
   in accordance with the    Eq.(\ref{Eq:FSS}) where we put ${ n_x=-1}$.

   General expressions for the amplitudes $K_{\sigma}$ are rather involved
   (see Appendix \ref{Sec:AppendixAndreevReflection}).
   However in the tunnelling limit they read
  { \begin{equation} \label{Eq:K}
   K_{\sigma }  = n_x F_{bcs}^* \frac{\sigma p_{\bar\sigma n} v_{Sn} }{2V^2},
   \end{equation}}
   where $F_{bcs} = \Delta/\sqrt{\omega^2 + |\Delta|^2}$ and
   $v_{Sn}=\sqrt{2 \mu_S/m_S}$.
   For the half-metal Eq.(\ref{Eq:K}) is valid with the imaginary momentum
   $p_{-n} = i \sqrt{2m_F (  h - \mu) +p_{||}^2}$.
   The other Nambu component of the anomalous function $\hat{ F} ( x,x^\prime) = (\check{G})_{12}$
  can be obtained from the general particle-hole symmetry
 {
  \begin{equation}\label{Eq:Symmetry}
  {\hat {\cal F}} (\omega) =\hat \sigma_y
   [{\hat {\tilde  {{\cal F }}}}  (\omega)]^* \hat\sigma_y  .
 \end{equation}
 }


Now having in hand the expression for the slowly-varying amplitude (\ref{Eq:FSS})
we can derive the boundary conditions for the components of
quasiclassical propagator.
Following Ref.~\onlinecite{Zaitsev1984}  we write them in the form
 \begin{equation} \label{Eq:ZaitsevAnsatz}
 \hat g_\sigma = g_\sigma \hat \tau_3 + f_\sigma \hat \tau_+ +  \tilde f_\sigma \hat \tau_- ,
 \end{equation}
  where $\hat\tau_{\pm } = (\hat\tau_x \pm i \hat\tau_y)/2$.
 The quasiclassical propagators can be obtained by taking the Fourier transform of the
 slow-varying exact GF components (\ref{Eq:FSS}) and then using the definition (\ref{Eq:DefinitionQuasiclassics}).
 In this way we obtain the propagators as functions of the momentum direction 
 ${\bm{ \hat p_\sigma} }$, which determines the quasiclassical
 trajectories in each of the spin subbands.
 We will use the notations
 $\tilde f_{\sigma,in(out)}$  for the "incoming" ("outgoing")
 trajectories with  $ \bm{ \hat p_\sigma}\bm n  < (>)0$ .

 Then  Eq.~(\ref{Eq:FSS}) yields the quasiclassical ESC propagators
at the interface $x=0$:
{\begin{eqnarray}
 \tilde f_{\sigma,in}(\omega>0) = 2i v_{\sigma n} K_{\sigma} S_{\sigma \perp}  ,
\label{tildef_in}
\end{eqnarray}}
where $v_{\sigma n} = p_{\sigma n}/m_F$ is the spin-dependent Fermi-velocity.
At the same time $\tilde f_{\sigma,out}(x=0,\omega>0)=0$.

The other anomalous Green's functions can be obtained from  $\tilde f_{\sigma}(x=0,\omega>0,\bm { \hat p_\sigma})$
according to the following symmetry relations \cite{Serene1983}:
$\tilde f_{\sigma}(x=0,\omega<0,\bm{ \hat p_\sigma}) = - \tilde f_{\sigma}(x=0,\omega>0, - \bm{ \hat p_\sigma})$ and
$f_\sigma(x=0,\omega, \bm{ \hat p_\sigma} ) = \tilde f_\sigma^*(x=0,-\omega, \bm{ \hat p_\sigma})$.
The normal part of the GF is given by $g_{\sigma,in(out)}=1$ due to the normalization condition.
Thus Eq.~(\ref{tildef_in}) gives the value of the ESC, generated by the magnetic inhomogeneity/SOC
at the FM/SC interface. To find this equal-spin GF in the ferromagnetic region
we solve in general the transport equation (\ref{Eq:Eilenberger}) with the boundary condition (\ref{tildef_in}).

{The advantage of the boundary conditions (\ref{tildef_in}) is that they give an explicit value of the Green's function at the ferromagnetic side of the interface. The price, which we have paid for it, is that, strictly speaking, they are only valid for an isolated interface, because the asymptotic conditions at the infinity were essentially used in the derivation. However, they can be safely applied to the dirty systems with more than one interface if the distances between the interfaces are large as compared to the mean free path. Below we are only interested in the dirty case.}

%
The boundary conditions to the Usadel equations can be obtained from Eq.~(\ref{tildef_in}) in a straightforward way.
As usual, one can show \cite{Kupriyanov1988} that in the isotropization region near the interface the matrix current is
  \begin{equation}
 \Bigl\langle \frac{v_{\sigma n}}{v_\sigma} \hat g_{\sigma,in}\Bigr \rangle_- -
 \Bigl \langle \frac{v_{\sigma n}}{v_\sigma}\hat g_{\sigma,out}\Bigr \rangle_+  =
 \frac{2l_\sigma }{3} \langle \hat g_\sigma \rangle \hat \partial_{\bm n} \langle \hat g_\sigma \rangle,
 \label{matrix_current}
 \end{equation}
 where $\langle ... \rangle_{-(+)}$ means the averaging over the part of the ferromagnet FS corresponding to
  $\bm{ p_\sigma} \bm n<(>)0$ and real
 values of $p_{\sigma n}$ and $p_{s n}$, $l_\sigma=v_\sigma \tau_\sigma$ is the mean free path
 and $\hat \partial_{\bm n}  =  \bm n \hat \partial_{\bm R}$.
  The boundary condition to the Usadel equation is obtained using (\ref{matrix_current}),
 taken at $x=0$ with $\hat g_{\sigma,in(out)}$ from Eq.~ (\ref{tildef_in}) and the symmetry relations discussed above:
 \begin{eqnarray}
 \frac{l_\sigma }{3} \langle \hat g_\sigma \rangle \hat \partial_{\bm n} \langle \hat g_\sigma \rangle =
 {\kappa_\sigma^* S_{\bar \sigma\perp}\hat\tau_+ - \kappa_\sigma S_{\sigma\perp} \hat\tau_-} ,
 \label{bc_usadel}
 \end{eqnarray}
 where $ \kappa_\sigma  = -i \langle K_{\sigma} v_{\sigma n}^2 / v_{\sigma} \rangle_{-}$.

 From the boundary condition (\ref{bc_usadel}) one can see that the generation of ESC is determined
  by the non-adiabatic spin-flipping terms near the boundary.
 In case if these terms are of the SOC origin, e.g. having the Rashba form the magnitude of ESC correlations is given by
 $S_{\sigma\perp}\sim \alpha /h$, where $\alpha$ is the SOC constant.
  Otherwise if it comes from the magnetic texture with the characteristic scale $\xi_\theta$ the estimation
 is $ S_{\sigma\perp}\sim 1 /(m_F \xi_\theta h)$.
 However as shown below, that smallness affects only the overall amplitude of the
  critical current in the generic  SC/FM/SC Josephson systems ,
   but not the spontaneous phase shift of the current-phase relation.
 Indeed the emergent gauge field $\bm Z$
 which drives the spontaneous supercurrents through strong ferromagnets does not contain any small parameter. Therefore
 the anomalous current at zero phase difference across the junction can be of the order of the critical one.

 \section{Spontaneous Josephson current through strong ferromagnets}
  {
  Having in hand the machinery of the generalized quasiclassical theory described above we
  can calculate Josephson current-phase relations for different systems with spin-dependent fields.
  Example of such a system with magnetic helix texture
  is shown in Fig.~\ref{Fig:Model}c .

 Here we work in the dirty limit using the linearized (with respect to the anomalous Green’s function) version of the 
 spin-less Usadel equations (\ref{Eq:UsadelEquation}) and boundary conditions (\ref {bc_usadel}).
  This simplification is adequate if the proximity effect at the SC/FM interface is weak, for example when the interface is low-transparent.
 The absolute value of the order parameter is assumed to be the same in the superconducting leads, while there is the phase difference
 $\chi$ between them. The electric current in the ferromagnetic interlayer can be calculated according to
 Eq.~(\ref{Eq:ChargeCurrentDiff}), which in the linearised form is reduced to
  \begin{equation}
  j=\frac{i \pi T e}{2} \sum \limits_{\omega,\sigma} \nu_\sigma D_\sigma
  \left( f_\sigma \partial_x \tilde f_\sigma -
  \tilde f_\sigma\partial_x f_\sigma -
  4i\sigma Z_x f_\sigma \tilde f_\sigma \right).
  \label{current_linear}
  \end{equation}
  The anomalous GF $f_\sigma$ and $\tilde f_\sigma$ should be calculated from the linearized version of the
  Usadel equation (\ref{Eq:UsadelEquation}):
  \begin{eqnarray}
  D_\sigma (\partial_x + 2 i \sigma Z_x)^2 f_\sigma -2|\omega|f_\sigma = 0 \nonumber \\
  D_\sigma (\partial_x - 2 i \sigma Z_x)^2 \tilde f_\sigma -2|\omega|\tilde f_\sigma = 0.
  \label{usadel_linear}
  \end{eqnarray}
  The solution of these equations takes the form:
  \begin{eqnarray}
  f_\sigma = \Bigl( C_{\sigma,+}e^{\lambda_\sigma x}+C_{\sigma,-}e^{-\lambda_\sigma x} \Bigr)e^{-2i \sigma Z_x x} \nonumber \\
  \tilde f_\sigma = \Bigl( \tilde C_{\sigma,+}e^{\lambda_\sigma x}+\tilde C_{\sigma,-}e^{-\lambda_\sigma x} \Bigr)e^{2i \sigma Z_x x}.
  \label{usadel_linear_solution}
  \end{eqnarray}
  The coefficients $C_{\sigma,\pm}$ and $\tilde C_{\sigma,\pm}$ are to be found from the boundary conditions (\ref{bc_usadel})
  taken at the S/F interfaces $x=\mp d/2$. The resulting expressions take the form:
  \begin{eqnarray}
  C_{\sigma,\pm}=-\frac{3{\rm sgn}\omega S_{\bar \sigma \perp}}
  {4 l_\sigma \lambda_\sigma \sinh[\lambda_\sigma d]}\times \nonumber \\
  \Bigl( \kappa_\sigma^{l*} e^{\mp \lambda_\sigma d/2-i\sigma Z_x d} -
  \kappa_\sigma^{r*} e^{\pm \lambda_\sigma d/2+i\sigma Z_x d} \Bigr),
  \label{C_values}
  \end{eqnarray}
  and
  \begin{eqnarray}
  \tilde C_{\sigma,\pm}=-\frac{3{\rm sgn}\omega S_{\sigma \perp}}
  {4 l_\sigma \lambda_\sigma \sinh[\lambda_\sigma d]}\times \nonumber \\
  \Bigl( \kappa_\sigma^l e^{\mp\lambda_\sigma d/2+i\sigma Z_x d} -
  \kappa_\sigma^r e^{\pm \lambda_\sigma d/2-i\sigma Z_x d} \Bigr),
  \label{tilde_C_values}
  \end{eqnarray}
  where
  $\lambda_\sigma=\sqrt{2|\omega|/D_\sigma}$.
  Substituting the anomalous GF from (\ref{usadel_linear_solution}) with the coefficients $C_{\sigma, \pm}$ and $\tilde C_{\sigma, \pm}$ from (\ref{C_values}) and (\ref{tilde_C_values}) into Eq.~(\ref{current_linear}) we obtain
   the general current-phase relation (CPR):
   }
  \begin{align} \label{josephson}
 & I (\chi) = \sum_{\sigma =\pm} I_\sigma \sin(\chi+2 \sigma  Z_x d) \\
 & \frac{e R_\sigma I_\sigma}{\pi}  =
 -
\frac{9 S_{\sigma\perp} S_{\bar \sigma\perp} } {l_\sigma^2} \sum \limits_{ \omega>0}  \frac{T |\kappa_\sigma|^2}{ \lambda_\sigma
\sinh(\lambda_\sigma d)},
\label{josephson1}
\end{align}
where $\chi$ is the Josephson phase difference
 $R_{\sigma} = 1/ (e^2 \nu_\sigma D_\sigma)$ is the spin-resolved resistivity and $\lambda_\sigma=\sqrt{2|\omega|/D_\sigma}$.

The spin gauge field $Z_x\neq 0$ and
finite spin splitting $D_+\neq D_-$ in
the CPR (\ref{josephson}) lead to the spontaneous  current at zero phase difference
known as the anomalous Josephson effect. The
ground state phase difference $\chi_0$ can be found from the zero-current condition $I (\chi  =\chi_0)=0$
\begin{eqnarray}
\tan \chi_0 = \frac{I_- - I_+}{I_- + I_+}\tan (2Z_x d).
\label{chi_0}
\end{eqnarray}

 The spontaneous phase shift of Josephson current has been obtained in several FM/SC systems
 \cite{Braude2007,Buzdin2008, Reynoso2008,Grein2009, Zazunov2009, Liu2010, Malshukov2010, Mironov2015, Brunetti2013,
Yokoyama2014, Kulagina2014,Nesterov2016, Konschelle2015, Moor2015, Moor2015a,Bobkova2016, Szombati2016, Silaev2017}.
Here we demonstrate that this effect is essential
only for the case of strong ferromagnets. When the ferromagnet is weak and
 treated within the usual  quasiclassical approximation,
 the difference between
 $I_-$ and  $I_+$  is neglected and the anomalous Josephson effect
 disappears.

 The spin gauge field $\bm Z$ is generated by the spin helix shown schematically in
 Fig.~ \ref{Fig:Model}. Recently the proximity effect in helical magnets has been observed
 experimentally\cite{DiBernardo2015,Satchell2017}.
 In this case the magnetization texture is described by
$ \bm h = h (\cos \alpha, \sin \alpha \cos \theta, \sin \alpha \sin \theta)$,
where we assume that the angle $\alpha$ is spatially independent and $\theta=\theta(x)$.
The spin rotation is given by
 $ \hat U = e^{-i\hat\sigma_x \theta/2} e^{-i\hat\sigma_z \alpha/2}
 e^{-i\hat\sigma_y \pi/4}$ yielding $Z_x=-\cos \alpha \partial_x \theta/2m_F$.
 The surface ESC-generating term in Eq.(\ref{Eq:FSS}) is provided by
$S_{\sigma\perp} = i \sigma n_x \sin \alpha \partial_x \theta /(2m_F h)  $ .

%
The general theory developed above describes the proximity effect in homogeneous ferromagnet
with a linear in momentum SOC \cite{Bergeret2013,Bergeret2014}.
For example let us consider the SC/FM/SC junction through the quasi-2D ferromagnet in $xz$ plane,  interfaces in $yz$
planes and the exchange field in the plane of the ferromagnet ${\bm h}\parallel {\bm z} $ .
In case of the Rashba SOC in the ferromagnetic region
this system is characterized by  the spin-dependent fields $M_{zx} = -M_{xz}= - \alpha/2$ which leads to
\begin{eqnarray}
Z_x = - m_F \alpha /2 ; \; S_{\sigma\perp} = 0,
\label{SO_Rashba}
\end{eqnarray}
while the Dresselhaus SOC yields $M_{zz}=-M_{xx} = \beta/2 $ and therefore
\begin{eqnarray}
Z_z = m_F \beta/2 ; \;\; S_{\sigma\perp} = -n_x \beta/2h .
\label{SO_Dress}
\end{eqnarray}
In each of these cases the Josephson CPR can be found
substituting the fields into the general Eqs.(\ref{josephson},\ref{josephson1}).
Since the ground state phase shift  $\chi_0$ is determined by the component $Z_x$
parallel to the Josephson current, for
 $\bm h \parallel {\bm z}$ we have
$\chi_0 \neq 0,\pi$  only for the Rashba SOC, but not for the Dresselhaus one.
This is natural, because in general case of a magneto-electric effect the spontaneous current and the magnetization are perpendicular to each other for the Rashba SOC \cite{Buzdin2008,Malshukov2008,Bobkova2017}, but they are parallel for the Dresselhaus SOC.
Therefore, in order to get the anomalous Josephson current for the Dresselhaus SOC,
${\bm h}$ is to have a component parallel to the current.

Comparing Eqs.(\ref{josephson1}) and (\ref{SO_Rashba},\ref{SO_Dress}) one can see that
 in the considered geometry the Dresselhaus SOC produce the long-range ESC even in the case
of the homogeneous magnet (while in general the both Rashba and Dresselhause SOC can produce ESC \cite{Bergeret2014}).
However their amplitudes are determined by the SOC constants which in general are rather small in metals.
Although the anomalous phase shift is also determined by SOC, it can become rather large for sufficiently long junction, i.e. when
$|Z_x d |\geq 1$.
Therefore even the weak SOC leads to the significant phase shifts of the CPR
although the overall critical current amplitude is rather small .
In reality however the long-range ESC can be generated by the magnetic inhomogeneity near the interface \cite{Eschrig2008}.
Let us consider the following model :
\begin{eqnarray}
\bm h = h(\sin \theta ,0,\cos \theta ), ~ U=e^{-i(\theta/2)\sigma_y},
\label{h_interface}
\end{eqnarray}
where $\theta=\theta (x)$ changes linearly in the region $\xi_\theta \ll d$ near the interfaces and $\theta =0$ in the bulk FM.
Neglect the effect of SOC we get
$S_{\sigma\perp} = -i \sigma n_x \partial_x \theta /(2m_F h) $ in the
CPR Eq.~(\ref{josephson}), where the spin gauge field $Z_x$ is determined by the SOC in the bulk FM.
For small exchange fields $h\ll\mu$ the anomalous phase shift $\chi_0$ is determined by the prefactor
$(I_+ - I_-)/(I_+ + I_-) \sim h/\mu $.  If we assume additionally (in general it can be not the case) that
$Z_x d \ll 1$, then $\chi_0 \sim  m_F (hd/\mu)\alpha $, in agreement with Ref.~\onlinecite{Buzdin2008}.

 \section{Conclusion}
To conclude, we have developed the generalized quasiclassical formalism to calculate the behaviour of long-range ESC in the ferromagnet. These correlations can be generated at the ferromagnetic/superconductor interface in the presence of either magnetization inhomogeneity
or SOC. The general conditions for ESC generation are derived in terms of the
SU(2) gauge fields. In the ferromagnetic material the behaviour of ESC is shown to be governed by the adiabatic spin gauge field which generates spontaneous superconducting
currents through strong FMs with magnetic texture or SOC.
 These results demonstrate that spontaneous superconducting currents exist as a robust and experimentally observable phenomenon in many superconducting/ferromagnetic systems studied in connection to the  superconducting spintronics \cite{Linder2015, Eschrig2015a}.

 \section{Acknowledgements}
 We thank A.S. Mel'nikov for stimulating discussions.
 This work was supported by the Academy of Finland.
 I.V.B. and A.M. B. acknowledge the support by the Program of Russian Academy of Sciences “Electron spin resonance, spin-dependent electron effects and spin technologies” and 
Russian-Greek Project N 2017-14-588-0007-011 “Experimental and theoretical studies of physical properties of low-dimensional quantum nanoelectronic systems”.

 \begin{widetext}
 \appendix

 \section {Spin-flip Andreev reflection coefficient} \label{Sec:AppendixAndreevReflection}

  Here we find the reflection coefficients of the electronic and hole-like
  waves in Eqs.(\ref{Eq:ur},\ref{Eq:vr}).
  For this purpose we use Eq.(\ref{Eq:u1Incidend}) as the incident wave coming from the ferromagnet to the FM/SC
  interface.
  To apply the boundary conditions (\ref{Eq:bc1-1}, \ref{Eq:bc1-3})
  we write solution in  the superconductor, as the superposition of two terms decaying at $x\to\infty$ :
  \begin{align}
  &\hat u_S = \hat C_1 e^{iq_1 x} + \hat C_2 e^{iq_2 x} \\
  &\hat v_S = \frac{i\Delta^*}{\omega + \Omega} \hat C_1 e^{iq_1 x} + \frac{i\Delta^*}{\omega - \Omega} \hat C_2 e^{iq_2 x}
  \end{align}
  where $q_1  = \sqrt{2m_S (\mu_S + i \Omega  )}$  and
  $q_2  =  - \sqrt{2m_S (\mu_S - i \Omega )}$ and $\Omega = \sqrt{\omega^2 + |\Delta|^2}$.
  Below we will neglect the imaginary part of $q_{1,2}$ and use $q_1 \approx /q_2\approx p_{Sn}$ where $p_{Sn}= \sqrt{2m_S \mu_S} $.

  First, let us find reflection coefficients without spin-flip, in zero order by small parameter $M^+_{x,y}p_{x,y} /h $.
  For this purpose we obtain the following system of equations :
  \begin{align}
 & a^{(0)}_+ + d^{(0)}_+ = c_1 + c_2
 \\
 & v_{+n} ( a^{(0)}_+ - d^{(0)}_+ ) = \alpha_0 c_1 - \alpha^*_0 c_2
 \\
 & b^{(0)}_+ =  i\Delta^* ( c_1/\omega_+ + c_2/\omega_- ) \\
 & v_{-n} b^{(0)}_+ = i\Delta^* (\alpha_0 c_1/\omega_+ - \alpha^*_0 c_2/\omega_- )
  \end{align}   	
  where
  \begin{align}
  \alpha_0 = v_{Sn} + 2i V \\
  \omega_\pm = \omega \pm \Omega.
  \end{align}

  The solutions are
  \begin{align} \label{Eq:a00}
  & a^{(0)}_+  =1/(i v_{+n})
  \\
  \label{Eq:d0}
  & d^{(0)}_+ = a^{(0)}_+
  \frac{ ( \alpha^*_0 + v_{+n} ) ( \alpha_0 - v_{-n} ) \omega_- -
  (\alpha^*_0 + v_{-n} ) ( \alpha_0 - v_{+n} ) \omega_+ }
  {(\alpha^*_0 + v_{-n} ) ( \alpha_0 + v_{+n} ) \omega_+ -
   (\alpha^*_0 - v_{+n}) ( \alpha_0 - v_{-n} ) \omega_- }
   \\
   \label{Eq:b0}
   & b^{(0)}_+ =
  a^{(0)}_+ \frac{2i\Delta^* v_{+n} v_{Sn} }{Y^*_+ }
  %
  \end{align}

  The spin-flip reflection amplitude $b_-$ can be found taking into account first-order corrections
  in $M^+_{x,y}p_{x,y} /h $ when matching the electron and hole waves at FM/SC boundary.
  In this way we get the linear system
  \begin{align}
  & \hat A^{(0)}_+ + \hat D^{(0)}_+ + \hat D_- = \hat C_1 + \hat C_2
  \\
  & v_{+n} (\hat A^{(0)}_+ - \hat D^{(0)}_+ ) - v_{-n} \hat D_- =
  \hat \alpha\hat C_1 -  \tilde{\hat\alpha} \hat C_2
  \\
  & \hat B^{(0)}_+ + \hat B_- = i\Delta(\hat C_1/\omega_+  + \hat C_2/\omega_- )
  \\
  & v_{-n} \hat B^{(0)}_+ + v_{+n} \hat B_- =
  i\Delta (\hat\alpha \hat C_1/\omega_+  - \tilde{\hat\alpha} \hat C_2/\omega_-),
  \end{align}
  where the coefficients
  $  \hat\alpha = \alpha_0 - M_{kx}\hat \sigma_k $ and
  $\tilde{\hat\alpha} = \alpha_0^* +  M_{kx}\hat \sigma_k $ take into account the correction to the boundary condition from
  the effective spin-orbital term.
  The spinors $ \hat A^{(0)}_+ $, $ \hat B^{(0)}_+ $, $ \hat D^{(0)}_+ $ are given by
  Eqs.(\ref{Eq:A-up-plus},\ref{Eq:D},\ref{Eq:B}) with the amplitudes (\ref{Eq:a00},\ref{Eq:d0},\ref{Eq:b0})
  and without the spin-flip terms
  %
  %
   \begin{align}
   \label{Eq:A0}
   & \hat A^{(0)}_+ = a^{(0)}_{+ >} e^{-ip_{+n} x^\prime} \left( 1, \; 0  \right)^T \\
   \label{Eq:D0}
   & \hat D^{(0)}_{+} = d^{(0)}_+ e^{-ip_{+n} x^\prime} \left( 1, \; 0 \right)^T \\
   \label{Eq:B0}
   & \hat B^{(0)}_+   = b^{(0)}_+ e^{-ip_{+n} x^\prime} \left( 1, \; 0 \right)^T .
   \end{align}
  The solution of this system reads
  \begin{equation} \label{Eq:Bdownarrow}
  \hat B_- = \frac{ i v_{Sn} \Delta^* ( v_\Sigma \hat A^0_+ -
  v_d \hat D^0_+ ) - \hat Z \hat B^0_+ }{ Y_+ },
  \end{equation}
  where
   \begin{align} \label{Eq:Z}
   & {
   \hat Z = Z_+ + 4i\Omega V M_{kx}\hat\sigma_k
   }
   \\ \label{Eq:Zuparrow}
   & Z_+ = \Omega ( |\alpha_0|^2 + v_{-n}^2  ) +
   2 \omega v_{Sn} v_{-n} =
   \frac{1}{2}( \omega_+ |\alpha_0 + v_{-n}|^2 -
   \omega_- |\alpha_0 - v_{-n}|^2 )
   \\ \label{Eq:Yuparrow}
   & Y_+ = \Omega ( |\alpha_0|^2 + v_{+n} v_{-n} + 2i v_d V ) + \omega v_{Sn} v_\Sigma =
   \frac{1}{2} [ \omega_+ (\alpha_0 + v_{-n})(\alpha_0^* + v_{+n})  - \omega_- (\alpha_0 - v_{+n})(\alpha_0^* - v_{-n}) ],
   \end{align}
    where $v_\Sigma = v_{+n} + v_{-n}$ and $v_d = v_{+n} - v_{-n}$.

    In this way we obtain the spin-flip Andreev reflection amplitude in the form
      ${ b_- = K_{x+} S_{+\perp} + K_{y+} S_{+\parallel}}$ with
 {{\begin{align} \label{Eq:ResFx}
& K_{x +} = -
  \frac{ i\Delta^* v_{Sn} p_{+n} ( a^{(0)} v_\Sigma + d_+^{(0)} v_d ) +
  b_+^{(0)}
  (
  Z_+  p_{-n} +  8h i\Omega V
  ) }
  {2Y_+ } \\ \label{Eq:ResFy}
  & K_{y +} = p_y
  \frac{ i\Delta^* v_{Sn} ( d_+^{(0)} v_d - a^{(0)} v_\Sigma ) -
  b_+^{(0)} Z_+ } {2Y_+ } .
  \end{align}}}

   { \bf (ii)}
  Let us now consider the other pair of spinors in Eq.(\ref{Eq:spinor}) $\hat u_2$, $\hat v_2$
  which determine the correlation function ${\tilde {\cal F}_-}$ .
  They can be obtained from $\hat u_1$, $\hat v_1$ by transforming the Hamiltonian
  $\check H \to \sigma_x \check H \sigma_x $ which flips the spin index $\sigma$.
  The reflected hole-like states are given by the same Eq.(\ref{Eq:B}), but this time we are
  interested in the wave $\hat B_{+} (x^\prime)$. This wave has the form
   (\ref{Eq:B}) with the amplitude ${ b_+ = K_{x-} S_{-\perp} + K_{y-} S_{-\parallel}}$, where
   {
  {\begin{align} \label{Eq:ResFx1}
  & K_{x -} =
  \frac{ i\Delta^* v_{Sn} p_{-n} ( a^{(0)} v_\Sigma - d_-^{(0)} v_d ) +  b_-^{(0)}
  (
  Z_- p_{+n} - 8hi\Omega V
  ) }
  {2Y_- }
  \\ \label{Eq:ResFy1}
  & K_{y-} =  p_y
  \frac{ i\Delta^* v_{Sn} \left(  d_-^{(0)} v_d + a^{(0)} v_\Sigma  \right) + b_-^{(0)} Z_-}
  {2Y_- } ,
  \end{align}}   }
  where the amplitudes $d_-^{(0)}$, $b_-^{(0)}$,  $a_-^{(0)}$
  and coefficients $Z_-$, $Y_-$
  are obtained from (\ref{Eq:a00},\ref{Eq:d0},\ref{Eq:b0},\ref{Eq:Bdownarrow},\ref{Eq:Z},
  \ref{Eq:Zuparrow},\ref{Eq:Yuparrow}) with $+$ changed by the $-$ and vice versa.

  { \bf  (iii) Limiting cases: }
  Consider large barrier { $V\gg v_S, v_\sigma , v_\sigma, h/p_\sigma $ } so that $\alpha \approx i V$
  and $Z_0\approx Y \approx \Omega V^2$. Then from Eqs.
  we get $d_+^{(0)} \approx - a_+^{(0)} $ and
  $b_+^{(0)} \approx  a_+^{(0)} i\Delta p_{Sn} p_{+n} /( 2\Omega V^2)$.

Substituting to the Eq.(\ref{Eq:ResFx}) and taking into account (\ref{Eq:a00}) we get
  \begin{align}
  K_{x+} = - \frac{v_{Sn} p_{-n} }{2V^2} \frac{\Delta^*}{\sqrt{\omega^2 + |\Delta|^2}}    \\
  K_{x-} =  \frac{v_{Sn} p_{+n} }{2V^2} \frac{\Delta^*}{\sqrt{\omega^2 + |\Delta|^2}}
  \end{align}
   {
   These expressions are valid for the half-metal when $h>\mu_F $ as well. In this case one should take
   $p_{-n} = i \sqrt{2m_F (  h - \mu_F) }$.
   }

\end{widetext}

 %


 \end{document}